\documentclass[twocolumn,prb,aps]{revtex4}
\usepackage{graphicx}
\usepackage{subfigure}
\usepackage{bm}
\usepackage{amsmath}
\begin{document} 
\title{Understanding highly excited states via parametric variations}
\author{Aravindan  Semparithi, 
Venkataraman Charulatha\footnote{Present address:
Department of Chemistry, University of California,
Berkeley, CA, USA 94720}, and Srihari Keshavamurthy}
\affiliation{Department of Chemistry, Indian Institute
of Technology, Kanpur, India 208 016}
\date{\today}
\begin{abstract}
Highly excited vibrational states of an isolated molecule
encode the vibrational energy flow pathways in the molecule.
Recent studies have had spectacular success in understanding the
nature of the excited states mainly due to the extensive studies
of the classical phase space structures and their bifurcations.
Such detailed classical-quantum correspondence studies are presently
limited to two or quasi two dimensional systems. 
One of the main reasons for such a constraint has to do with the
problem of visualization of relevant objects like surface of sections and
Wigner or Husimi distributions associated with an eigenstate.
This necessiates various alternative techniques which are more
algebraic than geometric in nature. 
In this work we introduce one such method based
on parametric variation of the eigenvalues of a Hamiltonian.
It is shown that the level velocities are correlated with the
phase space nature of the corresponding eigenstates. 
A semiclassical expression for the level velocities of
a single resonance Hamiltonian is derived which provides theoretical
support for the correlation.
We use the level velocities to dynamically assign the highly
excited states of a model spectroscopic Hamiltonian
in the mixed phase space regime.
The effect of bifurcations on the level velocities is briefly
discussed using a recently proposed spectroscopic Hamiltonian for
the HCP molecule.
\end{abstract}
\maketitle

\nopagebreak

\section{Introduction}

Determining the nature of the highly excited rovibrational states 
is important in the context of unraveling the 
intramolecular vibrational energy redistribution (IVR) pathways in 
molecules\cite{rev1,rev2,rev3,rev4,rev5,rev6,rev7}. 
At high energies the usual rigid rotor-harmonic oscillator
descriptions are inadequate due to the breakdown of the normal mode
approximation. The various normal modes get coupled leading to
complicated spectra and patterns. The underlying classical phase space
usually has a mixed character {\it i.e.,} regular regions interspersed
with the chaotic regions. 
The mixed nature of the phase space gives rise
to a variety 
of phenomena, classical\cite{uzer} and quantum\cite{tun}, 
which in turn lead
to complex spectral patterns in terms of the intensities and splittings.

Much of the recent work\cite{water,water1,ethy,chbr,dco,annu,jpc} on small 
polyatomic molecules has clearly
shown that the underlying classical dynamics plays a significant role
in determining the nature of the
highly excited states. In particular, the various classical phase space
structures like periodic orbits, resonance zones
and effects like bifurcations\cite{jpc,kell} leave their 
imprints on the eigenstates and
hence provide important clues to IVR. 
Almost all of the work has been
constrained to atmost two degrees of freedom systems. 
One of the main reasons for
such a constraint have to do with the fact that visualizing the classical
phase space structures, for that matter even the quantum states, 
is quite involved
for systems with more than two degrees of freedom.
Moreover, determining the relevant periodic orbits\cite{farant} 
and higher dimensional
objects like resonant 2-tori becomes 
difficult as the number of degrees of freedom
increases. It is obvious that
studying IVR in large molecules from such 
useful classical-quantum correspondence
viewpoints
would require one to come up with dimensionality independent techniques.
Although not the focus of the present work, it is important to note
that novel dynamical phenomena like Arnol'd diffusion\cite{arnold} 
and complex
instability\cite{cinst} 
arise in systems with dimensionality greater than two.
It remains to be seen wether such effects have important consequences
for the classical-quantum correspondence studies
of highly excited states and hence IVR.

In principle, once the detailed phase space
structure is at hand, 
the required information can be
extracted by correlating with a phase space representation
of the quantum states, like
the Husimi\cite{hus} or the Wigner\cite{wig} distributions.
However, given that a detailed classical phase space
study of even three degree of freedom 
systems is sparse\cite{nato,licht,ccm,mde,lask}
suggests that the straightforward approach is far from easy.
Ideally one would like a 
method that yields the required information
in a basis independent fashion
without the need to explicitly 
determine the classical periodic orbits, to visualize the surface of
sections and then comparing to the relevant phase space
distributions. 
The search for such a method, if it exists, seems to be
rather difficult at the present moment. 
In this work we attempt to provide
a first step towards such an important yet difficult goal.

The above mentioned factors clearly indicate that one needs
algebraic measures that would unambigously provide the information
on the nature of the eigenstates. 
To this end certain measures, time dependent and independent, have
been proposed in the literature. For instance, inverse participation
ratios\cite{ipr} (IPR) or dilution factors can yield information about the 
nature of the localization of the eigenstates. However IPR is
a basis dependent quantity and cannot provide information
about the invariant classical structures relevant to the eigenstate.
It is possible to extract useful information by computing the IPR
in different basis which are 
optimal and physically motivated\cite{rev5,kl}.
An explicit expression for the probability distribution
of the IPR has been derived very recently\cite{leit}.
From a time-dependent point of view valuable information can be
obtained by computing various 
survival probabilities\cite{rev5,woly,jf}.  
Insights regarding the
nature of the underlying classical phase space
have emerged from establishing the existence of correlated intermediate
timescale dynamics for rovibrational Hamiltonians\cite{rev5,woly}.
The role played by the various classical structures 
in determining the nature of the survival probabilities is 
well known\cite{hel}.
Recently\cite{kelld} it has been proposed that a dip in the level spacings
is a characteristic spectral pattern indicative of a seperatrix
in the underlying classical phase space. 
This is an important observation 
for fitting 
the experimental spectrum with
effective Hamiltonians since seperatrices are indicative
of qualitatively different dynamical behaviours coexisting in the system.

On the other hand,
when the system Hamiltonian is known,
a number of authors have focused attention on parametric variation
of the eigenvalues of the system\cite{ram,kay,sri}. 
In particular, plots of the energy
eigenvalues versus a perturbation parameter show very specific
patterns for resonant systems\cite{kay,sri}. 
It is also well known\cite{pech,lak,hak} 
that the parametric variation of the eigenvalues of
a system can be mapped onto a classical Hamiltonian system
with the eigenvalues corresponding to the positions of a set of
fictitious particles and the parameter playing the role of a psuedotime.
Within this mapping the slopes of the energy curves correspond
to level velocities\cite{hak}. 
Thus one associates a level velocity with every
eigenstate of the system. 
Significant progress has been made 
in studying the statistical properties
of the level velocities\cite{stvel} 
and the accelarations\cite{stcurv} (curvatures). 
In analogy with the random matrix theory (RMT) results for eigenvalue
statistics, the level velocity and curvatures exhibit universal
behaviours in the RMT limit\cite{hak}. For example, in RMT 
the level velocities
are found to be gaussian distributed and a relation\cite{fyd} has been
established between the IPR distribution and the level velocity
distribution.
Accordingly, deviations from the RMT limit result in nonuniversal
distributions signifying 
localization of the eigenstates\cite{tak,nick,arul,kct}.
The preceeding discussion suggests that it might be useful to
explore the possibility of using the level velocities to understand the
highly excited states of an isolated molecule. 

The purpose of this paper is to 
show that the level velocity of an eigenstate is
strongly correlated with the phase space nature of the eigenstate.
It will be demonstrated that the magnitude and relative signs of the
level velocities give a direct indication of the important classical
structures that influence the corresponding eigenstates.
Theoretical arguments for the correlation
and numerical support are provided in the
next section for
classically integrable, single resonance systems.
In section III the level velocity ``spectrum" of
a nonintegrable, multiresonant system is studied.
Combined with an earlier work on the fingerprints of resonances
on the eigenlevel dynamics it is possible to have an approximate
dynamical assignment of the eigenstates when the underlying
phase space is mixed.
We conclude with a discussion of the major advantages and disadvantages
of our approach and briefly discuss the effects of bifurcation
on the level velocities.


\section{Theoretical background}
In this section we introduce the level velocity spectrum and provide
arguments for the observed correlation between level velocities and
the phase space nature of the corresponding eigenstates. 
The arguments presented in this section are strictly applicable only
to single resonant Hamiltonians. Nevertheless,
we expect the arguments to be 
valid for the multiresonant systems exhibiting mixed phase space
dynamics.

Consider the following $n:m$ resonant Hamiltonian:
\begin{eqnarray}
\widehat{H}(\tau)&=&\sum_{j=1,2}^{} 
(\omega_{j} N_{j} + \alpha_{jj} N_{j}^{2}) \nonumber \\
&+& \tau \left[(a_{1}^{\dagger})^{m} (a_{2})^{n} +  {\rm h.c} \right]
\end{eqnarray}
with $N_{j} = (n_{j} + 1/2)$ and $(a_{j}, a_{j}^{\dagger})$ denoting
the standard destruction and creation operators for the j$^{\rm th}$ mode.
The eigenvalues and eigenstates of the Hamiltonian are denoted by
$E(\tau)$ and $|\alpha(\tau)\rangle$ respectively.
The form of the diagonal part of $\widehat{H}(\tau)$ is not restricted to the
choice made above and can include higher order anharmonicities as well.
The form of the Hamiltonian is that of an effective or spectroscopic
Hamiltonian and $\widehat{H}_{0}$ corresponds to the Dunham expansion. 
Defining the polyad operator\cite{poly} 
$\widehat{P} = (n/m) \widehat{n}_{1} +\widehat{n}_{2}$
it is easy to see that $[\widehat{H},\widehat{P}] = 0$. Thus $\widehat{H}$ is
block diagonal with each block labeled by the polyad number $P$
and all the eigenstates can be 
assigned in terms of two quantum numbers as
$|\alpha(\tau) \rangle = |P,\nu(\tau) \rangle$. 
The quantum number $\nu = 0,1,\ldots$ 
is an excitation index\cite{water,sri} whose range
is related to the strength of the resonant coupling $\tau$.
The eigenstates for a given $P$ are linear combinations of the
zeroth order basis functions $|{\bf n} \rangle \equiv |n_{1},n_{2} \rangle$
with $(n/m) n_{1} + n_{2} = P$. Hence we can write $|\alpha(\tau) \rangle = 
\sum_{{\bf n} \in P} c_{{\bf n} \alpha}(\tau) |{\bf n} \rangle$ with
$c_{{\bf n} \alpha}(\tau) = \langle {\bf n} | \alpha(\tau) \rangle$.

The level velocity associated with an eigenstate $|\alpha(\tau)\rangle$
is obtained using the Hellman-Feynman theorem and given by:
\begin{equation}
\dot{x}_{\alpha}(\tau;P) \equiv \frac{\partial E(\tau)}{\partial \tau} 
= \langle \alpha(\tau) |
\hat{V}_{n:m} | \alpha(\tau) \rangle
\end{equation}
with $\hat{V}_{n:m}$ being the perturbation.
It is possible to write the level velocity in terms of the expansion
coefficients $c_{{\bf n} \alpha}(\tau)$ as
\begin{equation}
\dot{x}_{\alpha}(\tau;P) = 
\sum_{{\bf n} \neq {\bf n}'}^{{\bf n},{\bf n}' \in P} 
c^{*}_{{\bf n} \alpha}(\tau)
c_{{\bf n}' \alpha}(\tau) V_{{\bf n} {\bf n}'}
\end{equation}
where the matrix element $V_{{\bf n} {\bf n}'} =
\langle {\bf n} | \widehat{V}_{n:m} | {\bf n}' \rangle$.
At this stage we summarize the salient features of $\dot{x}_{\alpha}(\tau;P)$
for single resonant systems. 
\begin{enumerate}
\item The average of the level velocities for states belonging to a given
polyad $P$ vanishes irrespective of the
coupling strength $\tau$. This can be easily seen since
\begin{eqnarray}
\sum_{\nu \in P} \dot{x}_{\alpha}(\tau;P) &=& 
\sum_{{\bf n} \neq {\bf n}'}^{{\bf n},{\bf n}' \in P} 
V_{{\bf n} {\bf n}'} \sum_{\nu \in P}
\langle {\bf n}'| P;\nu \rangle \langle P;\nu | {\bf n} \rangle \nonumber \\
&=& \sum_{{\bf n} \neq {\bf n}'}^{{\bf n},{\bf n}' \in P}
V_{{\bf n} {\bf n}'} 
\langle {\bf n}'| {\bf n} \rangle = 0
\end{eqnarray}
This fact gives rise to the ``fan-like" structures in the
energy correlation diagram\cite{water,sri} 
with the variation parameter being $\tau$.
\item The level velocities exhibit linear parametric motion\cite{sri} 
beyond a 
certain critical coupling strength $\tau_{c}$. The
critical coupling can
be approximately determined via a Chirikov\cite{chiri} analysis. 
It has been argued previously\cite{sri}
that the asymptotic velocities for a $1:1$ state $|P;\nu \rangle$
are given by $(P - 2 \nu)$. 
In general, one does not expect the 
resonant coupling strength in a 
molecule,
to be close to $\tau_{c}$. Consequently the velocities, except
for states localized at or near the center of the resonance zone,
do show nonlinear
parametric variation. 
However, as an example we mention that the fermi resonance
coupling strength in the spectroscopic effective
Hamiltonian of the CHF$_{3}$ molecule does come close to the critical
value\cite{chf3}.
For a general $n:m$ resonant 
Hamiltonian it was conjectured\cite{sri}, and numerically
confirmed, that the velocity of the state $|P;0 \rangle$ scales with
the polyad as $P^{(m+n)/2}$.
\end{enumerate}

It is clear that through the velocity one is studying the response of the
eigenstate to a specific perturbation. It is not surprising that
states localized in or near the specific resonance zone would
have the largest response. 
That the response is strongly correlated
to the phase space nature of the eigenstate is interesting
and we now provide
plausible reasons for such an observation.
We note that there has been an earlier work wherein similar correlations were
observed\cite{jort}.

To start with note that the matrix element
\begin{equation}
\langle {\bf n} | (a_{1}^{\dagger})^{m} 
(a_{2})^{n} |{\bf n}' \rangle  
= V_{nm}(n_{1}',n_{2}') 
\delta_{n_{1},n_{1}'+m} \delta_{n_{2},n_{2}'-n}
\end{equation}
yields an expression for the level velocity in the following
convenient form:
\begin{equation}
\dot{x}_{\alpha}(\tau;P) = 2 \sum_{n_{1}'=0}^{P} 
V_{nm}(n_{1}';P) c_{n_{1}'+m,\alpha}^{*}(\tau) c_{n_{1}',\alpha}(\tau)
\label{delos}
\end{equation}
where we have supressed the second index on the coefficients.
From a classical-quantum correspondence perspective it would be ideal
to have a semiclassical expression for $\dot{x}_{\alpha}(\tau;P)$.
The shifted overlap like structure 
in the above expression for $\dot{x}_{\alpha}(\tau;P)$ is strongly
reminiscient of the Wigner function\cite{wig}.
To make the correspondence explicit the level velocity is written as:
\begin{subequations}
\begin{eqnarray}
\dot{x}_{\alpha}(\tau;P) &=& \langle \alpha |\widehat{V}_{n:m} |\alpha \rangle 
\\
&=& {\rm Tr} \left[ \widehat{V}_{n:m} |\alpha \rangle \langle \alpha | \right] 
\\
&=& \frac{1}{2 \pi \hbar} \int dI \, d\phi \, V_{W}(I,\phi) 
W_{\alpha}(I,\phi)
\end{eqnarray}
\end{subequations}
where $W_{\alpha}(I,\phi)$ is the Wigner function associated with the 
state $|\alpha \rangle$ and $V_{W}(I,\phi)$ 
is the Weyl symbol\cite{weyl,weyl1,weyl2}
of the operator $\widehat{V}_{n:m}$ defined by
\begin{equation}
V_{W}(I,\phi) = \int d\phi' e^{-i I \phi'/\hbar} \langle \phi + \phi'/2 |
\widehat{V}_{n:m} | \phi - \phi'/2 \rangle
\end{equation}
Note that the Wigner function is the Weyl symbol of the density operator.
Although the system is two dimensional it is sufficient to focus on
an effective one dimensional Hamiltonian due to the existence of
the polyad. Thus $(I,\phi)$ are action-angle
variables pertinent to the classical limit of
the reduced zeroth order Hamiltonian (cf. appendix).
In particular, all of the expressions in this section
are for a given polyad $P$.
This already suggests the role played by classical phase space
structures since it is known\cite{berry} that for integrable systems the Wigner
function condenses onto the invariant torus corresponding to
the eigenstate.  

In order to determine the Weyl symbol $V_{W}(I,\phi)$ consider
\begin{eqnarray}
\int  &d\phi'& e^{-i I \phi'/\hbar} \langle \phi + \phi'/2 |
(a_{1}^{\dagger})^{m} (a_{2})^{n} |  \phi - \phi'/2 \rangle \nonumber \\
&=& \sum_{n_{1},n_{1}'} \frac{1}{2 \pi} 
\int  d\phi' e^{-i I \phi'/\hbar} e^{i (n_{1}-n_{1}') \phi}
e^{i (n_{1}+n_{1}') \phi'/2} \nonumber \\
&\times& V_{nm}(n_{1}';P) \delta_{n_{1},n_{1}'+m} 
\nonumber \\
&=& \sum_{n_{1}'} V_{nm}(n_{1}';P)\, e^{i m \phi}\,
\delta \left[\frac{I}{\hbar} - (n_{1}' + \frac{m}{2}) \right]
\end{eqnarray}
Since the Weyl symbol of a 
hermitian operator is real\cite{weyl1,weyl2}, $V_{W}(I,\phi)$
is given by:
\begin{equation}
V_{W}(I,\phi) = 2 V_{nm}(I/\hbar-m/2;P) \cos m\phi
\end{equation}
The Wigner function is now approximated\cite{weyl} in order to 
elucidate the role of the fixed points.
First the wavefunction is written in its primitive semiclassical form
\begin{equation}
\psi_{\alpha}(I) \approx \sum_{b} A_{b \alpha}(I)
e^{-i S_{b\alpha}(I)/\hbar}
\end{equation}
where the index $b$ denotes the branch. 
The phase and amplitude are given by
\begin{subequations}
\begin{eqnarray}
S_{b\alpha}(I) &=& \int_{}^{I} \varphi_{b \alpha}(I')\, dI' + 
\frac{\pi \hbar}{2} \mu_{b \alpha} \\
A_{b \alpha} &=& \left|\frac{\partial H_{cl}(I,\phi)}
{\partial \phi}\right|_{\phi = \varphi_{b \alpha}(I)}^{-1/2}
\end{eqnarray}
\end{subequations}
The Maslov index\cite{weyl} is denoted by $\mu_{b \alpha}$
and $H_{cl}(\varphi_{b \alpha}(I),I) = E_{\alpha}$.
Next the Wigner function is approximated as 
\begin{eqnarray}
W_{\alpha}(I,\phi) &=& \int d \sigma e^{i \sigma \phi/\hbar}
\psi_{\alpha}(I+\sigma/2) \psi_{\alpha}^{*}(I-\sigma/2) \nonumber \\
&\approx& 2 \pi \hbar \sum_{b} 
|A_{b \alpha}(I)|^{2} \delta[\phi-\varphi_{b \alpha}(I)]
\end{eqnarray}
by expanding
the amplitude and phase terms
to first order. The intrabranch terms have been neglected in the 
approximation.
The above result for the
Wigner function highlights it as the classical density 
on an invariant torus in the corresponding
phase space. 
Combining the approximate Wigner function with
the Weyl symbol for $\widehat{V}_{n:m}$  
one arrives at the central result of this section-
a approximate semiclassical expression for the level velocity associated
with an eigenstate $|\alpha \rangle$ as
\begin{equation}
\dot{x}_{\alpha}(\tau;P) \approx 2 \sum_{b} \int_{0}^{P_{c}}dI \, 
\frac{1}{|\dot{I}|} V(I;P_{c}) \cos[m \varphi_{b \alpha}(I)]
\label{semvel}
\end{equation}
where $P_{c}$ is the classical analog of $P$ and the leading
order $\hbar$ term of $V_{nm}(I/\hbar-m/2;P)$ is denoted by
$V(I;P_{c})$.
The above semiclassical expression emphasizes the fact that
fixed points of the classical Hamiltonian {\it i.e.,} $\dot{I} = 0$
play an important role in the level velocities. 
Moreover for a single resonant Hamiltonian the phase term in the
level velocity expression is $\pm 1$
depending on the nature of the fixed
point. Thus, relative signs of the
level velocities are important.
In the appendix further insight is provided 
on the polyad scaling of the velocities at such fixed
points by a classical analysis of the function
$V(I;P_{c})$.
Based on the above analysis it is expected that the level velocities
will be large in magnitude for states localized around the
various fixed points.
A similar expression for the level velocity can be derived\cite{unpub} 
independently starting
from the Eq.~(\ref{delos}) and following 
an approach based on psuedodifferential
equations\cite{delos}.

To provide support for the theoretical arguments presented above consider
an effective 2-mode Hamiltonian $\widehat{H} = \widehat{H}_{0}
+ \tau \widehat{V}_{1:1}$
which is $1:1$ resonant with
the parameters $\omega_{1} = 1, \omega_{2} = 0.8, \alpha_{11} = -0.03,
\alpha_{22} = -0.02$, and $\tau = 0.01$.
In order to determine the phase space nature of the various eigenstates
the Husimi distributions are computed. Although one could compute
the Wigner functions associated with the states the Husimi functions
are preferred due to their smooth nature\cite{takahas}.

\begin{figure}
\includegraphics*[width=3in,height=3in]{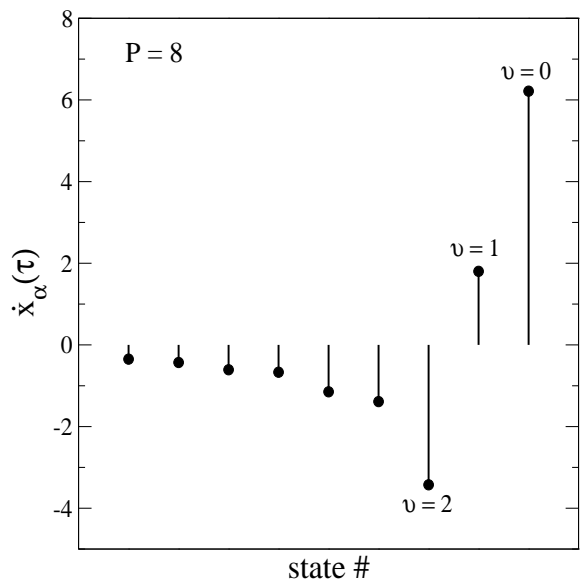}
\caption{Level velocity spectrum associated with the states
belonging to the polyad $P = 8$ for the single $1:1$ resonant
Hamiltonian.}
\label{fig 1}
\end{figure}

\begin{figure}
\includegraphics*[width=3in,height=3in]{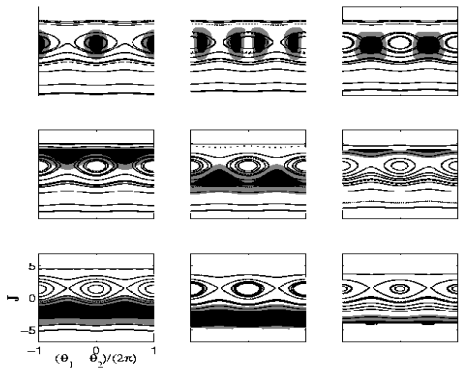}
\caption{Husimi distributions associated with the states
belonging to the polyad $P = 8$ for the single $1:1$ resonant
Hamiltonian. For comparison the Husimis are superimposed
on the surface of sections. The top three panels show the resonant
states with $\nu = 0,1,2$.}
\label{fig 2}
\end{figure}

In Fig.~(\ref{fig 1}) we show the level velocities for the states belonging to
polyad $P = 8$. The three states which lie inside the $1:1$
resonance zone are indicated by the label $\nu$. It is easy to
see that the coupling is below the critical threshold since the
velocity of the $|8;0 \rangle$ state is less than the asymptotic
value of $8$. 
In Fig.~(\ref{fig 2}) the Husimis associated with the states are
shown superimposed on the appropriate Poincar\'e surface of
sections. For the sake of clarity only the contours with the
maximum values are shown. 
The state $|8;0 \rangle$ is localized on the stable
periodic orbit {\it i.e.,} at the center of the $1:1$ island.
As one proceeds to the states with higher $\nu$ the Husimi 
maximum moves out of the island center and
for the state $|8;2 \rangle$ eventually
localizes on the unstable periodic orbit {\it i.e.,} around the
seperatrix.
It is important to note that the state $|8;0 \rangle$ has a large positive
velocity and the state $|8;2 \rangle$ has a large negative velocity. 
This observation is consistent with the results of the theoretical
analysis.
The rest of the states in $P = 8$ are either local modes or distorted
local modes as evident from the Husimis in Fig.~(\ref{fig 2}).

\begin{figure}
\includegraphics*[width=3in,height=3in]{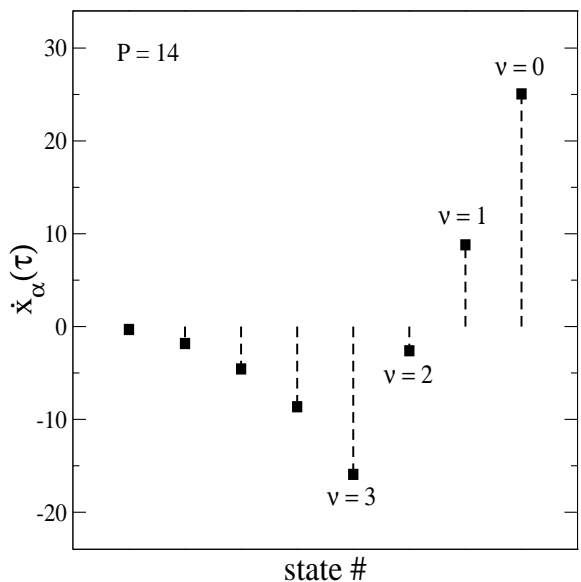}
\caption{Level velocity spectrum associated with the states
belonging to the polyad $P = 14$ for the single $2:1$ resonant
Hamiltonian.}
\label{fig 3}
\end{figure}

\begin{figure}
\includegraphics*[width=3in,height=3in]{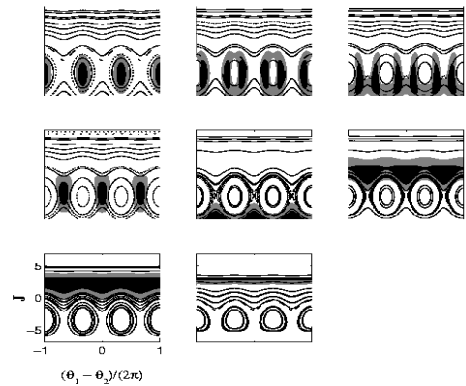}
\caption{Husimi distributions associated with the states
belonging to the polyad $P = 14$ for the single $2:1$ resonant
Hamiltonian. The top three and the middle left panels correspond
to the resonant states with $\nu = 0,1,2,3$ respectively.}
\label{fig 4}
\end{figure}

In order to further check the correlation between level velocities
and the phase space nature of the corresponding states the velocities
and Husimis are computed for an integrable $2:1$
system $\widehat{H} = \widehat{H}_{0}
+ \tau \widehat{V}_{2:1}$
with the coupling strength $\tau = 0.01$.
The level velocities for the states belonging to the $2:1$ polyad
$P = 14$ are shown in Fig.~(\ref{fig 3}). 
Again it is clear that the coupling is
below the critical coupling and the level velocities show a 
characteristic pattern. The state $|14;0 \rangle$ is 
expected to localize around
the $2:1$ stable periodic orbit whereas the state $|14;3 \rangle$
should be localized around the unstable periodic orbit. The 
husimis shown in Fig.~(\ref{fig 4}) essentially confirm the expectations.
In both the integrable cases, for coupling strengths near or above
the critical value, the level velocities are large positive for
the state in the center of the resonance island and equally large but
negative for the seperatrix state.
Thus the relative signs of the level velocities are crucial and provide
information on the fixed points that dictate the phase space nature
of the eigenstates. The absolute sign is clearly unimportant since a
change in the sign of the coupling would interchange the signs
of the velocities.
In addition the state localized about the stable periodic orbit
exhibits linear parametric motion and the velocity is closer
to the classical value, computed using Eq.~(\ref{clvel}), as 
compared to the state localized about the
unstable periodic orbit.

The results of this section clearly indicate that the level velocities
associated with the eigenstates are correlated to the structures in
the classical phase space. In the following 
sections additional numerical examples
are provided which highlight the correlation and the potential for using
the level velocities to assign highly excited states
of multiresonant Hamiltonians.

\section{Application to nonintegrable systems}

The states of single resonant Hamiltonians studied in the previous section
can be assigned in various ways. Complications due to bifurcations
can be sorted out\cite{kell,annu} with a 
detailed study of the classical-quantum
correspondence for the system. However with the addition of another
independent resonance the polyad constant of the motion is destroyed and
assignments of the states becomes a nontrivial task. 
The underlying classical phase space can exhibit a rich variety of
dynamics ranging from the near-integrable to 
the mixed to full chaos.
An important observation that has emerged from 
the various studies\cite{water,ethy,chbr,dco,annu,jpc}
is that classical phase space structures can form the basis
for a dynamical assignment of the various eigenstates of the system.
The advantage of such a dynamical assignment is that it 
is based on invariant structures and directly
reflects the relevant molecular motions in some energy range of interest. 
More importantly, such dynamical assignments are possible
over the entire range of the phase space dynamics.
For instance, a completely
chaotic phase space at the outset seems to
preclude any meaningful assignment of
the states. 
Even for such hard chaotic
systems it is wellknown that the eigenstates can be localized due to various
quantum and classical effects\cite{scar,cantor1,cantor2,bsep,diff}. 
A prime example is the scarring\cite{scar} of
the eigenstates by the periodic orbits which has been studied
in great detail\cite{scar2} over the years. 
In this case the relevant periodic
orbits provide an avenue to dynamically assign the eigenstates.

From the perspective of studying IVR in isolated molecules
a key observation is that the underlying phase space overwhelmingly belongs
to the class of mixed dynamics.
The fact that the effective spectroscopic Hamiltonians 
used to gain insights into
IVR have the structure of a hierarchical local random matrix\cite{rev5}
is largely responsible for the mixed nature of the phase space. 
Dynamical assignment of the highly excited eigenstates 
in such mixed phase space regimes is
important and highly relevant. 
A significant bottleneck to such a classical-quantum correspondence based
approach arises for systems with dimensionality greater than two.
Any dynamical assignment of the eigenstates necessarily involves
finding the important classical structures in the
phase space and correlating them with the eigenstates
or some phase space representation of the eigenstates.
Already for a three degree of freedom system the Poincar\'{e} surface of
section\cite{licht} is four dimensional and so is the computed
Wigner or Husimi distribution of the eigenstates.
This leads to diffculties in visually correlating the states with appropriate
classical objects. One way to overcome this problem is to have
algebraic measures and in this regard the results of the previous section
suggest the level velocities as one possible candidate.

In order to assess the utility of the level velocities in providing a
dynamical assignment of the eigenstates a model two resonance
effective 
Hamiltonian $\widehat{H} = \widehat{H}_{0} + \tau_{11} \widehat{V}_{1:1}
+ \tau_{21} \widehat{V}_{2:1}$ is chosen. 
The parameters of the zeroth order Hamiltonian are as given in the previous
section. The resonant coupling strengths 
are chosen to be $\tau_{11} = \tau_{21} = 0.01$.
The Hamiltonian is nonintegrable and the energy range $[8,9]$ is chosen
as an example since the phase space exhibits mixed dynamics.
The level velocities $\dot{x}_{\alpha}(\tau_{11})$ and
$\dot{x}_{\alpha}(\tau_{21})$ are computed for some of the
eigenstates in the selected energy range.
Note that in the nonintegrable case the velocities can be expressed as:
\begin{eqnarray}
\dot{x}_{\alpha}(\tau) &=& 2 \sum_{P} \sum_{n_{1}=0}^{P}
V_{nm}(n_{1};P) c^{*}_{n_{1}+m,\alpha}(\tau) c_{n_{1},\alpha}(\tau) \\
&\equiv& 2 \sum_{P} \dot{x}_{\alpha}(\tau;P)
\end{eqnarray}
In this work the $\dot{x}_{\alpha}(\tau;P)$ will be called as
the partial level velocities.
The level velocities are scaled as 
\begin{equation}
\widetilde{\dot{x}}_{\alpha}(\tau) = \frac{1}{\sigma_{E}}
(\dot{x}_{\alpha}(\tau) - \langle \dot{x}_{\alpha}(\tau)
\rangle)
\end{equation}
where $\sigma_{E}$ is the variance of the velocities
and $\langle \dot{x}_{\alpha}\rangle$ is the average of the 
level velocities.
This is done since the velocities scale differently with the polyad $P$
and we would like to compare the two velocities for a given
eigenstate.
The scaled velocities are dimensionless,
zero centered and have unit variance.

\begin{figure}
\includegraphics*[width=3in,height=3in]{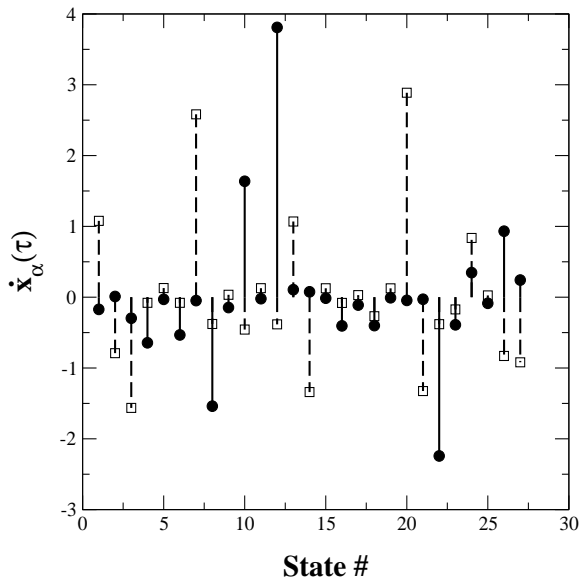}
\caption{Scaled level velocity spectrum associated with the states
in the energy range $[8,9]$ for the multiresonant 
Hamiltonian. The solid lines with filled circles correspond to the 
$1:1$ case and the dashed lines with open squares correspond to
the $2:1$ case.}
\label{fig 5}
\end{figure}

In Fig.~(\ref{fig 5}) we show the scaled level velocity spectrum 
for selected eigenstates in the energy range $[8,9]$.
In Table~\ref{table 1} the state energies, state IPRs in various basis, the
scaled level velocities and the assignments are provided.
The IPR of a state $|\alpha \rangle$ in a basis ${|b \rangle}$ can
be obtained as
\begin{equation}
L_{b}(\alpha) = \sum_{b} | \langle b | \alpha \rangle |^{4}
\end{equation}
In Table~\ref{table 1}, $L_{0}, L_{11}$, and $L_{21}$ denote the IPRs in the
zeroth order number basis, the $1:1$ basis and the $2:1$ basis
respectively.
The standard interpretation of the IPR is that a low value signifies
extensive mixing in the basis of choice whereas a high value
indicates very little mixing of the basis states.
The scaled velocity spectrum in Fig.~(\ref{fig 5}) clearly shows 
different classes of states 
for which a brief description is provided below.

\begin{table}
\caption{Assignment
of selected states for the nonintegrable
system.}
\label{table 1}
\begin{ruledtabular}
\begin{tabular}{cccccccc}
State & $E$ & $L_{0}$ & $L_{11}$ & $L_{21}$
& $\widetilde{\dot{x}}_{11}$ & $\widetilde{\dot{x}}_{21}$ &
Assignment\footnote{$s$, $u$, $l$, and $dl$ denote 
stable, unstable, local, and distorted local respectively.
States involved in avoided
crossings are marked with a $*$. See the text for details.} \\ \colrule
68 & 8.038 & 0.35 & 0.39 & 0.89 & -0.17 & 1.08 & $P_{21}-1$ \\
69 & 8.044 & 0.84 & 0.85 & 0.98 & 0.01 & -0.79 & $dl_{21}$ \\
70 & 8.164 & 0.22 & 0.28 & 0.79 & -0.30 & -1.56 & $u,P_{21}$ \\
71 & 8.189 & 0.74 & 0.95 & 0.77 & -0.65 & -0.08 & $dl_{11}$ \\
72 & 8.203 & 0.99 & 1.00 & 0.99 & -0.03 & 0.13 & $l$ \\
74 & 8.253 & 0.25 & 0.28 & 0.71 & -0.53 & -0.08 & * \\
75 & 8.277 & 0.35 & 0.37 & 0.85 & -0.05 & 2.58 & $s,P_{21}-1$ \\
77 & 8.313 & 0.21 & 0.75 & 0.26 & -1.54 & -0.38 & $u,P_{11}$ \\
79 & 8.338 & 0.94 & 0.99 & 0.95 & -0.15 & 0.03 & $l$ \\
80 & 8.367 & 0.33 & 0.70 & 0.41 & 1.64 & -0.46 & $P_{11}$ \\
82 & 8.423 & 0.99 & 1.00 & 0.99 & -0.02 & 0.13 & $l$ \\
85 & 8.460 & 0.36 & 0.71 & 0.43 & 3.81 & -0.38 & $s,P_{11}$ \\
86 & 8.473 & 0.21 & 0.23 & 0.81 & 0.11 & 1.07 & $P_{21}$ \\
88 & 8.514 & 0.38 & 0.39 & 0.93 & 0.08 & -1.34 & $u,P_{21}+1$ \\
89 & 8.583 & 0.99 & 1.00 & 0.99 & -0.01 & 0.13 & $l$ \\
91 & 8.650 & 0.83 & 0.97 & 0.85 & -0.41 & -0.08 & $dl_{11}$ \\
92 & 8.678 & 0.95 & 0.99 & 0.96 & -0.11 & 0.03 & $l$ \\
93 & 8.682 & 0.22 & 0.25 & 0.76 & -0.40 & -0.27 & * \\
94 & 8.683 & 0.99 & 1.00 & 0.99 & -0.01 & 0.13 & $l$ \\
96 & 8.719 & 0.33 & 0.35 & 0.85 & -0.05 & 2.89 & $s,P_{21}$ \\
98 & 8.763 & 0.48 & 0.47 & 0.96 & -0.03 & -1.33 & $u,P_{21}+2$ \\
99 & 8.848 & 0.24 & 0.58 & 0.27 & -2.24 & -0.38 & $u,P_{11}+1$ \\
100 & 8.869 & 0.18 & 0.20 & 0.27 & -0.39 & -0.17 & * \\
101 & 8.898 & 0.18 & 0.19 & 0.53 & 0.35 & 0.84 & *,$2:1$ \\
102 & 8.958 & 0.96 & 0.99 & 0.97 & -0.09 & 0.03 & $l$ \\
103 & 8.967 & 0.15 & 0.39 & 0.21 & 0.93 & -0.83 & *,$1:1,P_{11}+1$ \\
104 & 8.972 & 0.37 & 0.40 & 0.51 & 0.24 & -0.92 & *,$2:1$  \\
\end{tabular}
\end{ruledtabular}
\end{table}

\subsection{Local mode states}
These states (72,79,82,89,92,94,102)
are neither influenced by the $1:1$ perturbation nor the
$2:1$ perturbation and hence nonresonant. The scaled velocities
$\widetilde{\dot{x}}_{\alpha}(\tau_{11})$ and
$\widetilde{\dot{x}}_{\alpha}(\tau_{21})$ are both very small
indicating the local nature of these states.
Note also that the IPRs are close to one in every basis for these
set of states.
As an example, the husimi of state 94 superimposed on the surface of section
is shown in Fig.~(\ref{fig 6}a).
The zeroth order quantum numbers $(n_{1},n_{2})$ are good quantum numbers
for the local mode states.
In Table~\ref{table 1} such states are denoted by '$l$'.

\subsection{Distorted local mode states}
\begin{figure}
\includegraphics*[width=3in,height=3in]{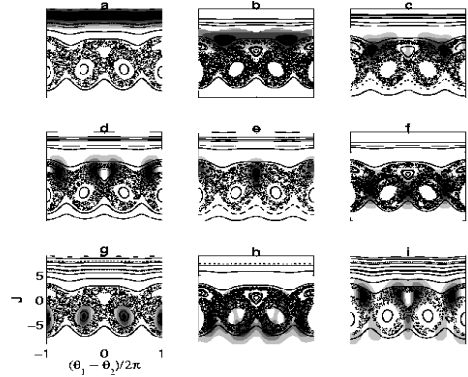}
\caption{Husimis for selected states superimposed on the corresponding
surface of sections. See the text for
details.}
\label{fig 6}
\end{figure}

As suggested by the name the distorted local mode states (69,71,91)
are states that are lying close to either one of the resonance
zones. 
Thus the husimi distribution of such states are distorted but topologically
nonresonant. 
One such example is shown in Fig.~(\ref{fig 6}b) for the state 71.
The level velocities clearly indicate the particular resonance zone to
which the states are in proximity. 
This is precisely the reason that the state 71 is labeled as '$dl_{11}$'.
The IPR values are also high and essentially agree with the assignments
from the level velocity spectrum.
The zeroth order quantum numbers are not so good a label for these
distorted states. It is also possible to think of these states as a kind
of 'borderline' states and hence a slight increase in the appropriate
resonance coupling strength will render them resonant.

\subsection{Resonant states}
The resonant states (68,70,75,77,80,85,86,88,96,98,99)
show up in the velocity spectrum in Fig.~(\ref{fig 5}) as states
with large positive or negative velocities.
In accordance with our discussions in the previous sections these
states are localized in one or the other resonance zone.
Moreover states with large positive velocity are expected to be
localized about the corresponding stable periodic orbit. On the
other hand states with large negative velocities are expected to be
localized about the corresponding seperatrix region.
For example state 85 has a large positive 
$\widetilde{\dot{x}}_{\alpha}(\tau_{11})$ and comparitively negligible
$\widetilde{\dot{x}}_{\alpha}(\tau_{21})$. Consequently this state is
anticipated to be localized exclusively in the $1:1$ resonance zone
and the Husimi distribution to be localized about the $1:1$ resonance
island in the underlying phase space.
That this is indeed the case is confirmed in Fig.~(\ref{fig 6}e).
Thus state 85 is labeled by '$s$' indicating localization about the
stable periodic orbit. In Table~\ref{table 1}
 we have also appended a label $P_{11}$
suggesting that the particular $1:1$ polyad $P_{11}$ is almost
a good quantum number to assign the state.
Similarly state 96 is localized about the $2:1$ island according to
the level velocities and Fig.~(\ref{fig 6}g) supports the assignment. An
approximate polyad $P_{21}$ is also attached to this state.

\begin{figure}
\includegraphics*[width=3in,height=3in]{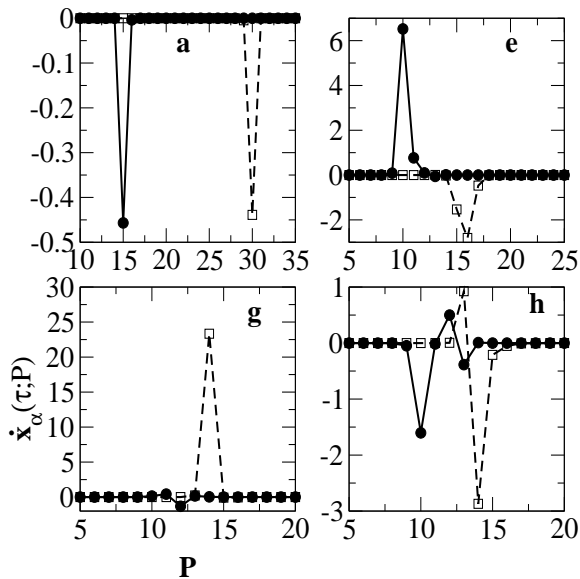}
\caption{Unscaled partial level velocities for
selected states as a function of the polyad.
The labeling of the figures corresponds to that in 
Fig.~(\ref{fig 6}).
The solid lines with filled circles correspond to the
$1:1$ case ($P = n_{1} + n_{2}$)
and the dashed lines with open squares correspond to
the $2:1$ case ($P = 2 n_{1} + n_{2}$).}
\label{fig 7}
\end{figure}

In order to complete the assignment the approximate
polyads $P_{11}$ and $P_{21}$ need to be determined.
To this end
it is useful to examine the partial level velocities
$\dot{x}_{\alpha}(\tau;P)$ as a function of $P$ for the 
corresponding states. 
$P$ is chosen appropriately for the particular resonance.
Thus, $P = n_{1} +n_{2}$ for the $1:1$ resonance and
$P = 2 n_{1} + n_{2}$ for the $2:1$ resonance.
It is important to note that this information is obtained
enroute to computing the level velocity of a state and does not
require a seperate computation.
In Fig.~(\ref{fig 7}e,g) the partial velocities
are shown for states 85 and 96 respectively. 
It is clear from the figure that $P_{11} = 10$ and
$P_{21} = 14$. A similar analysis can be done on all the states
and as an example the partial velocities are shown in Fig.~(\ref{fig 7}a)
for the state 94. One can immediately assign this local mode
state as $(n_{1},n_{2}) = (0,15)$

Once the resonant states with large positive velocities have been 
assigned the rest of the resonant states can be easily assigned
as shown in Table~\ref{table 1}. 
Two important properties of the level velocities
guide the assignments. The first property is that for positive coupling
constants the states localized about the stable periodic orbits
with some approximate polyad $P$
are energetically higher than the states localized about the
unstable periodic orbits belonging to $P$. 
Exceptions can occur due to bifurcations and this is discussed
in the following section. Secondly, 
velocity of states with
some approximate $P$ scale differently with $P$ for the $1:1$ and the
$2:1$ resonances. Thus states belonging to the $2:1$ resonance are
seperated more as compared to the states belonging to the
$1:1$ resonance. A Chirikov\cite{chiri} like estimate is also an indicator
of the number of states bound by the resonance.
Moreover, studying the partial velocities 
helps in confirming the assignments.
In Figs.~(\ref{fig 6}c,f) we show the Husimi distributions for the states
77 and 70 respectively. The Husimis are localized in the chaotic
regions corresponding to the integrable limit seperatricies
and lend support to the assignments in Table~\ref{table 1} where the label
'$u$' is used to denote such states.

\begin{figure}
\includegraphics*[width=2.5in,height=5.0in]{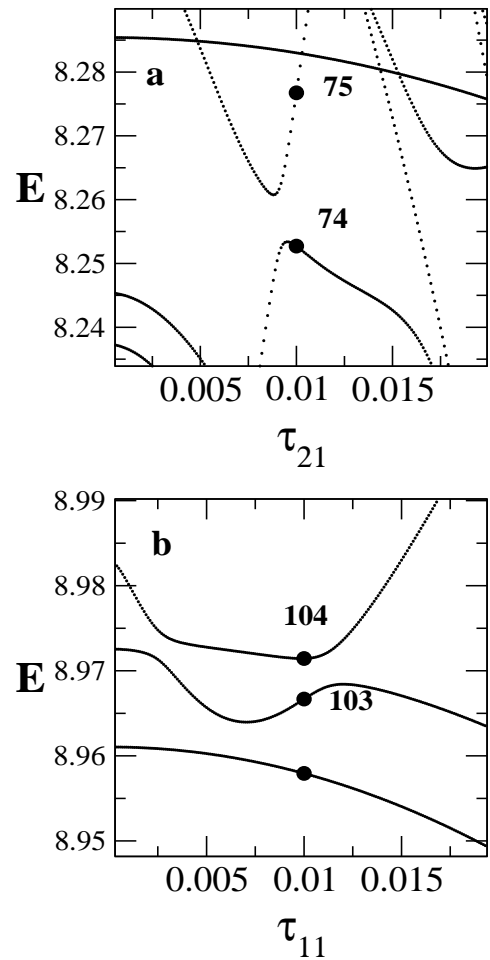}
\caption{(a) Sharp avoided crossing 
between states 74 and 75 as a function of the
$2:1$ coupling parameter $\tau_{21}$. (b) Broad avoided crossing between states
103 and 104 as a function of the $1:1$ coupling parameter $\tau_{11}$.
States are analysed in this work for $\tau_{11} = \tau_{21} = 0.01$.}
\label{fig 8}
\end{figure}

\subsection{Mixed states}
There are a few states (74,93,100,103) which cannot be assigned in
a straightforward fashion. As an example state 74 according to
the IPR values should be predominantly a $2:1$ resonant state. However
the level velocities show the opposite trend. 
In Fig.~(\ref{fig 6}h) we show the
Husimi for state 74 superimposed on the surface of section. It is
clear that the state is delocalized over both
the chaotic as well as the regular 
regions of the phase space as compared to the
other states in Fig.~(\ref{fig 6}). The Husimi, however, is peaked in the
$2:1$ regions of the phase space in agreement with the IPR values.
These observations are symptomatic of avoided crossings since at
an avoided crossing regular and chaotic states can mix. 
Indeed on examining the variation of the eigenvalues with
the $2:1$ coupling parameter $\tau_{21}$, shown in Fig.~(\ref{fig 8}a), 
it is seen that the state 74
is involved in an avoided crossing with the regular state 75. 
The important observation is
that the state 74 is located close to the center of the avoided crossing.
This explains the low value for the velocity
$\widetilde{\dot{x}}_{\alpha}(\tau_{21})$ and the apparent difficulty
in the assignment of the state 74. 
The partial velocities for the state 74 shown in Fig.~(\ref{fig 7}h) 
clearly demonstrate the delocalized nature of the state.
The assignment of the state 93 is complicated due to similar reasons.
The states 100 and 101 are also close to
being involved in an avoided crossing. In this case the velocities
can be used to provide a nominal assignment.

On the other hand states 103 and 104 are again
close to being involved in an avoided crossing
as shown in Fig.~(\ref{fig 8}b). 
Note that the parameter being varied in Fig.~(\ref{fig 8}b) is the $1:1$
coupling strength $\tau_{11}$.
The husimi distribution for the state 103 is
shown in Fig.~(\ref{fig 6}i) and looks topologically very much like the
case in Fig.~(\ref{fig 6}d) except for the peak in the $2:1$ region.
Again the level velocities can be used to nominally assign the state
103 as a $1:1$ state belonging to the polyad $P_{11}+1$.
Thus the mixed states arising in this energy range are primarily
due to two state avoided crossings. 

\section{Discussion and summary}
The main focus of the present work is to come up with techniques
which would allow a dynamical assignment of highly excited states
of multiresonant, multidimensional ($> 2$) Hamiltonians. 
It is well known that an intimate knowledge of the various
classical phase space structures is the first important step
towards such dynamical assignments. However 
obtaining a detailed picture, akin to the two dimensional systems,
of the phase space structure for systems with three or larger degrees
of freedom is diffcult. In this paper we have
argued that the phase space nature of an eigenstate is encoded
in the manner that it responds to variation of
specific parameters of the Hamiltonian. In other words, the level velocity
associated with an eigenstate is strongly correlated to the 
phase space nature of the eigenstate. Theoretical arguments for
the observed correlation have been
provided for single resonant systems.
In a way the current work begins to provide a basis for understanding
and utilizing the usual energy correlation diagrams.
We have shown the utility of the level velocities in a
dynamical 
assignment of the highly excited states of a 2-mode multiresonant Hamiltonian.
It is important to point out here that our approach does
not require explicit determination of the periodic orbits for the
system. It is equally relevant to mention that the level velocity
approach is independent of the dimensionality of the system
and is manifestly basis invariant. Thus the
current work is a first step in coming up with algebraic measures
facilitating the study of
the classical-quantum correspondence for multidimensional systems.

\begin{figure}
\includegraphics*[width=3.5in,height=3.5in]{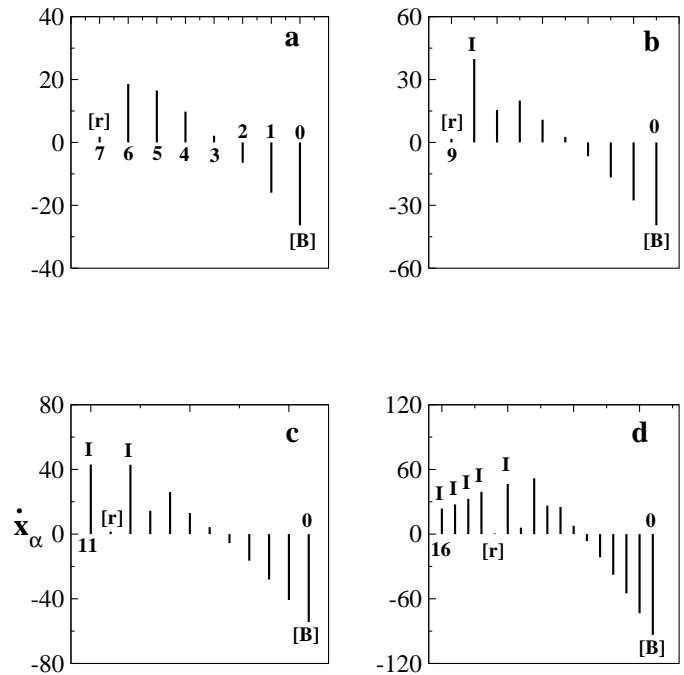}
\caption{Level velocity spectra for the HCP molecule for
the various CP stretch-bend polyads $P$.
$[r]$ and $[B]$ are stable periodic orbits.
(a) $P = 14$ (b) $P = 18$. Notice the perturbation of the positive part
of the level velocities. $I$ labels the isomerization state.
(c) $P = 22$ (d) $P = 32$.} 
\label{fig 9}
\end{figure}

The method proposed in this paper for the dynamical assignment of
highly excited states, however, has a few weakpoints. Firstly,
it is quite possible for the level velocity of a particular 
eigenstate to be close to zero. For example, for coupling strengths
near the critical value, the $1:1$ state $|6;3 \rangle$ will
exhibit zero velocity. Assigning such states solely on the basis of
level velocities will be problematic. Secondly, as demonstrated in
the last section, states involved in an avoided crossing can pose
problems for the assignment as well. This is not surprising since such
states are difficult to assign by any of the existing methods.
Nevertheless, taking linear combinations of the two states and
assigning the demixed states is one possible solution.
In both instances it is still possible to gain insights into the states
by examining other measures like the IPRs in conjunction with the
level velocities. 

One of the important issues that we have not addressed 
has to do with the effect of the various 
bifurcations\cite{kell,annu,jpc} on the 
level velocity spectrum. 
In this work we have explicitly shown the important role played by the
fixed points in the classical phase space as far as the level
velocities are concerned. Bifurcations\cite{licht,weyl} are associated with
creation or destruction of fixed points and hence signal qualitative
change in the dynamics.
Thus, it is clear that the
level velocities would be quite sensitive to bifurcations. 
Work is in progress in our group to study the fingerprints of bifurcations
on the level velocity spectrum in a sytematic fashion.
However, we conclude this paper by showing some preliminary work
on the HCP molecule in order to gauge the utility of the
level velocities.
We have chosen HCP as an example due to the fact that 
the system has been 
investigated experimentally\cite{hcpexpt} and thoroughly studied from
the classical-quantum perspective\cite{annu,jpc}.
The effective spectroscopic Hamiltonian\cite{hcp2000} 
for HCP involves a
$1:2$ Fermi resonant coupling between the bend and the CP stretch.
Thus there exists a polyad\cite{note} $P = v_{2} + 2 v_{3}$ with $v_{2}$ and
$v_{3}$ refering to the bend and the CP stretch modes respectively.
The states can be assigned in terms of the vibrational angular momentum,
the number of quanta in the CH stretch $(v_{1})$, and
the polyad $P$. Further for every given $P$ there are $P/2 + 1$
levels indexed by $i = 0$ at the top of the polyad and
$i = P/2$ at the bottom of the polyad.
It has been established\cite{hcp2000}
that for $v_{1} = 0$ and up to $P = 14$ the
states evolve smoothly from being pure CP stretch at the
bottom of the polyads to
being approximately pure bend at the top of the polyads.
However starting at $P = 18$ a new family of states, called isomerization
states, come into existence due to a saddle node bifurcation of
the periodic orbits. Above $P = 22$ the isomerization states
are almost pure bend and closely follow the minimum energy path
leading from HCP to CPH\cite{annu}. 
Keeping this information in mind
in Fig.~(\ref{fig 9}) we show the unscaled level velocity spectrum 
corresponding to polyads
$P = 14,18,22$, and $32$ for the HCP molecule.
It is crucial to note that 
for the effective spectroscopic Hamiltonian fit the Fermi resonant coupling
is negative\cite{hcp2000}. 
We thus anticipate a large negative velocity for the state
localized around the stable periodic orbit.
It is also known that for low polyad values there are two
stable periodic orbits denoted by $[r]$ and $[B]$. 
The velocity spectrum for $P= 14$ 
shown in Fig.~(\ref{fig 9}a) clearly indicates that
the $i=0$ state is localized around the $[B]$ periodic orbit. From our
theoretical analysis and the appendix we expect the state associated with
the $[r]$ periodic orbit to have a very small velocity which is
confirmed in Fig.~(\ref{fig 9}a) for state $i=7$. The velocity spectrum
suggests that most of the states are resonant and indeed the classical
estimate $(-28.51)$ (cf. Eq.~(\ref{clvel}))
for the velocity of the state $i=0$ is very close to the
computed value $(-26.36)$.
In Fig.~(\ref{fig 9}b) the velocity spectrum 
is shown for $P = 18$. For this case we
clearly see a pertubation in the spectrum in the positive
velocity region and the resulting isomerization state is labeled as $I$.
The negative velocity region is unperturbed.
The perturbation is quite 
prominent in Figs.~(\ref{fig 9}c,d) corresponding to
the $P = 22$ and the $P = 32$ cases. 
In every case the velocity of the $[B]$ state is close to the
classical estimate. Given that the negative portion of the velocity
spectrum is unperturbed we conclude that the bifurcation is creating
a stable periodic orbit $(SN)$
and an unstable periodic orbit $(\overline{SN})$
at $\psi = \pi/2$ in the reduced phase space.
Thus the isomerization states form a sequence from being localized about
the $SN$ periodic orbit to the $\overline{SN}$ periodic orbit.
In this instance as the saddle node bifurcation creates the stable/unstable
pair at the same angle $\psi=\pi/2$ the level velocities for states
localized about both the periodic orbits are expected to be large
positive.
The caricature of the phase space just described agrees well with
the earlier studies\cite{hcp2000,jpc}.
It is clear that the level velocity spectrum is correlated with
the phase space structure and capable of detecting nontrivial
changes in the phase space due to the bifurcations.

This work is a first step towards developing tools suitable for 
studying classical-quantum correspondence in multidimensional systems.
The systems studied in this paper are certainly encouraging
and warrant further study of the level velocities and
different measures based on the level velocities.
One such sensitive measure has been proposed recently\cite{kct} 
which correlates the
level velocities with the eigenstate resolved spectrum of the molecule
in order to gain insights into the IVR process.
Currently we are exploring 
the utility of the level velocities in assigning the states of
three degree of freedom multiresonant
Hamiltonians which are effectively two dimensional due to the existence of
a superpolyad number. 
Spectroscopic Hamiltonians have been determined for
many such effectively two dimensional
systems like H$_{2}$O\cite{bagg}, CHBrClF\cite{beil},
and DCO\cite{temp}.
Preliminary studies\cite{prelim} on such systems 
indicate that the level velocity spectrum is well suited for the
purpose of dynamical assignment of the eigenstates.
Studies are also underway of true three degree of freedom systems
in order to test the approach and for 
insights into the highly excited states of such systems from the classical-
quantum correspondence perspectives.

\section{Acknowledgements}
SK gratefully acknowledges the Department of
Science and Technology and the Council for Scientific and Industrial
Research, India for financial support.

\section*{Appendix: Classical insight into the level velocities}

In this appendix we provide a classical argument for the scaling
of the level velocities with the polyad $P$ for a general single
resonance Hamiltonian.

The $n:m$ resonant Hamiltonian 
corresponding to the classical limit of the
quantum Hamiltonian is:
\begin{equation}
H({\bf I},{\bm \theta}) = H_{0}({\bf I}) + 2 \tau I_{1}^{m/2} I_{2}^{n/2}
\cos(m\theta_{1}-n\theta_{2})
\end{equation}
The classical limit Hamiltonian can be reduced to an effective one
dimensional one by using the following generating function
\begin{equation}
F_{2} = I\left(\theta_{1} - \frac{n}{m}\theta_{2}\right) + P_{c} \theta_{2}
\end{equation}
The reduced Hamiltonian is obtained as
\begin{equation}
H(I,\phi;P_{c}) = H_{0}(I;P_{c}) + 2 \tau V(I;P_{c})
\cos m \phi
\end{equation}
with
\begin{equation}
V(I;P_{c}) = I^{m/2} \left(P_{c}
- \frac{n}{m} I\right)^{n/2}
\end{equation}
The action $P_{c}$ is a
constant of the motion 
since the Hamiltonian is ignorable in the angle conjugate 
to $P_{c}$.
$P_{c} = n/m I_{1} +I_{2}$ is the classical analog of the polyad number.

The fixed points in the $(I,\phi)$ plane correspond to periodic
orbits in the full phase space. Using $\dot{I} = 0 = \dot{\phi}$
we obtain the fixed points as $\bar{\phi}= 0,
\pm \pi/m$. The solution $\bar{I} = 0$ is discarded since it is unphysical and
the other root of $\bar{I}$ is of no interest here since that corresponds to
all the excitation in the first mode.
Using the value of the angle $\phi$ at the fixed points the
corresponding actions are obtained by a solution to the algebraic equation:
\begin{equation}
\frac{\partial H_{0}}{\partial I} \pm 2 \tau
g(I;P_{c}) = 0
\end{equation}
where
\begin{equation}
g(I;P_{c}) = \frac{dV(I;P_{c})}{d I} 
\end{equation}
This nonlinear equation is solved numerically to obtain the periodic
orbits of the system.

In this appendix, however, attention is focused on
the analysis of the function
\begin{equation}
v(I,\phi;P_{c}) = \frac{\partial H}{\partial \tau} = 2 
V(I;P_{c}) \cos m \phi
\end{equation}
which occurs in the
semiclassical approximation Eq.~(\ref{semvel})
for the quantum level velocities. 
Interest in this function arises from exploring the maximum and minimum
values that this function can take at the fixed points in the 
$(I,\phi)$ space. Inspecting the $I$ and $\phi$ derivative of
$v(I,\phi;P_{c})$ 
\begin{subequations}
\begin{eqnarray}
\frac{\partial}{\partial \phi} v(I,\phi;P_{c}) &=& -2 m
V(I;P_{c}) \sin m \phi \\
\frac{\partial}{\partial I} v(I,\phi;P_{c}) &=& 2 g(I;P_{c}) \cos m \phi
\end{eqnarray}
\end{subequations}
it is clear that the critical points in the
$\phi$ variable are identical with those of the fixed points.
The critical points arising from the $I$ variable are
$\bar{I} = 0,(m/n)P_{c},m^{2} P_{c}/n(m+n)$. 
Note that demanding $g(I;P_{c}) = 0$ is tantamount to being on the center of
the resonance line and hence the stable periodic orbit.
The first two critical values give rise
to zero velocities and correspond to the ``edge" states in a given polyad
multiplet.
The third critical point gives rise to the velocities:
\begin{subequations}
\begin{eqnarray}
v(\bar{I}^{(3)},0;P_{c}) &=& - v(\bar{I}^{(3)},\pi/m;P_{c}) \\
&=& 2 \left[\frac{\mu^{n-m}}{(1+\mu)^{n+m}} \right]^{1/2} P_{c}^{(m+n)/2}
\label{clvel}
\end{eqnarray}
\end{subequations}
where, $\mu = n/m$. From the above one immediately sees that the
classical object takes the same absolute value at the stable and
unstable fixed points. Moreover, the classical analog exhibits
the identical scaling with the polyad $P_{c}$ as seen in exact quantum studies
done previously. In fact it is seen in the quantum studies that the
resonant eigenstate $|P;\nu\rangle$ with $\nu = 0$ {\it i.e.,} state
localized on the stable fixed point shows very good agreement with
the above classical estimate.

\end{document}